\documentclass[twoside,english,british]{article}
\usepackage[T1]{fontenc}
\usepackage[utf8]{inputenc}
\usepackage[pdftex,a4paper]{geometry}
\usepackage{amsmath}
\usepackage{amssymb}
\usepackage{esint}

\makeatletter

\DeclareFontEncoding{LGR}{}{}

\newcommand{\lyxmathsym}[1]{\ifmmode\begingroup\def\b@ld{bold}
  \text{\ifx\math@version\b@ld\bfseries\fi#1}\endgroup\else#1\fi}

\providecommand{\tabularnewline}{\\}

\usepackage{rotating}
\usepackage{multirow}
\usepackage{colortbl}
\definecolor{lightgray}{gray}{0.8}
\definecolor{orange}{rgb}{1,0.5,0}
\DeclareUnicodeCharacter{0391}{\ensuremath{\Alpha}}
\DeclareUnicodeCharacter{03B1}{\ensuremath{\alpha}}
\DeclareUnicodeCharacter{0392}{\ensuremath{\Beta}}
\DeclareUnicodeCharacter{03B2}{\ensuremath{\beta}}
\DeclareUnicodeCharacter{0393}{\ensuremath{\Gamma}}
\DeclareUnicodeCharacter{03B3}{\ensuremath{\gamma}}
\DeclareUnicodeCharacter{0394}{\ensuremath{\Delta}}
\DeclareUnicodeCharacter{03B4}{\ensuremath{\delta}}
\DeclareUnicodeCharacter{0395}{\ensuremath{\Epsilon}}
\DeclareUnicodeCharacter{03B5}{\ensuremath{\epsilon}}
\DeclareUnicodeCharacter{0396}{\ensuremath{\Zeta}}
\DeclareUnicodeCharacter{03B6}{\ensuremath{\zeta}}
\DeclareUnicodeCharacter{0397}{\ensuremath{\Eta}}
\DeclareUnicodeCharacter{03B7}{\ensuremath{\eta}}
\DeclareUnicodeCharacter{0398}{\ensuremath{\Theta}}
\DeclareUnicodeCharacter{03B8}{\ensuremath{\theta}}
\DeclareUnicodeCharacter{0399}{\ensuremath{\Iota}}
\DeclareUnicodeCharacter{03B9}{\ensuremath{\iota}}
\DeclareUnicodeCharacter{039A}{\ensuremath{\Kappa}}
\DeclareUnicodeCharacter{03BA}{\ensuremath{\kappa}}
\DeclareUnicodeCharacter{039B}{\ensuremath{\Lambda}}
\DeclareUnicodeCharacter{03BB}{\ensuremath{\lambda}}
\DeclareUnicodeCharacter{039C}{\ensuremath{\Mu}}
\DeclareUnicodeCharacter{03BC}{\ensuremath{\mu}}
\DeclareUnicodeCharacter{039D}{\ensuremath{\Nu}}
\DeclareUnicodeCharacter{03BD}{\ensuremath{\nu}}
\DeclareUnicodeCharacter{039E}{\ensuremath{\Xi}}
\DeclareUnicodeCharacter{03BE}{\ensuremath{\xi}}
\DeclareUnicodeCharacter{039F}{\ensuremath{\Omicron}}
\DeclareUnicodeCharacter{03BF}{\ensuremath{\omicron}}
\DeclareUnicodeCharacter{03A0}{\ensuremath{\Pi}}
\DeclareUnicodeCharacter{03C0}{\ensuremath{\pi}}
\DeclareUnicodeCharacter{03A1}{\ensuremath{\Rho}}
\DeclareUnicodeCharacter{03C1}{\ensuremath{\rho}}
\DeclareUnicodeCharacter{03A3}{\ensuremath{\Sigma}}
\DeclareUnicodeCharacter{03C3}{\ensuremath{\sigma}}
\DeclareUnicodeCharacter{03A4}{\ensuremath{\Tau}}
\DeclareUnicodeCharacter{03C4}{\ensuremath{\tau}}
\DeclareUnicodeCharacter{03A5}{\ensuremath{\Upsilon}}
\DeclareUnicodeCharacter{03C5}{\ensuremath{\upsilon}}
\DeclareUnicodeCharacter{03A6}{\ensuremath{\Phi}}
\DeclareUnicodeCharacter{03C6}{\ensuremath{\varphi}}
\DeclareUnicodeCharacter{03A7}{\ensuremath{\Chi}}
\DeclareUnicodeCharacter{03C7}{\ensuremath{\chi}}
\DeclareUnicodeCharacter{03A8}{\ensuremath{\Psi}}
\DeclareUnicodeCharacter{03C8}{\ensuremath{\psi}}
\DeclareUnicodeCharacter{03A9}{\ensuremath{\Omega}}
\DeclareUnicodeCharacter{03C9}{\ensuremath{\omega}}
\DeclareUnicodeCharacter{00A0}{~}
\DeclareUnicodeCharacter{00AC}{\ensuremath{\neg}}
\DeclareUnicodeCharacter{00B1}{\ensuremath{\pm}}
\DeclareUnicodeCharacter{00D7}{\ensuremath{\times}}
\DeclareUnicodeCharacter{00F7}{\ensuremath{\div}}
\DeclareUnicodeCharacter{2026}{\ldots}
\DeclareUnicodeCharacter{207A}{\ensuremath{^{+}}}
\DeclareUnicodeCharacter{207B}{\ensuremath{^{-}}}
\DeclareUnicodeCharacter{2020}{\ensuremath{\dagger}}
\DeclareUnicodeCharacter{2021}{\ensuremath{\ddagger}}
\DeclareUnicodeCharacter{2212}{\ensuremath{-}}
\DeclareUnicodeCharacter{2192}{\ensuremath{\rightarrow}}

\makeatother

\usepackage{babel}

\begin{document}

\title{Exploring Two Novel Features for EEG-based Brain-Computer Interfaces:
Multifractal Cumulants and Predictive Complexity}

\author{Nicolas Brodu <nicolas.brodu@univ-rennes1.fr>%
\thanks{Nicolas is with the Laboratoire du Traitement du Signal et de l’Image
(LTSI) at University of Rennes 1, Campus de Beaulieu, Bât 22, 35042
Rennes Cedex - France.%
},\\
Fabien Lotte <fprlotte@i2r.a-star.edu.sg>%
\thanks{Fabien is with the Institute for Infocomm Research (I2R), Signal Processing
Department, Brain-Computer Interface Laboratory, at 1 Fusionopolis
way, 138632, Singapore.%
},\\
Anatole Lécuyer <anatole.lecuyer@irisa.fr>%
\thanks{Anatole is with the Institut National de Recherche en Informatique
et en Automatique (INRIA), at Campus universitaire de Beaulieu, Avenue
du Général Leclerc, 35042 Rennes Cedex - France.%
}}

\date{August 2010}
\maketitle
\begin{abstract}
In this paper, we introduce two new features for the design of electroencephalography
(EEG) based Brain-Computer Interfaces (BCI): one feature based on
multifractal cumulants, and one feature based on the predictive complexity
of the EEG time series. The multifractal cumulants feature measures
the signal regularity, while the predictive complexity measures the
difficulty to predict the future of the signal based on its past,
hence a degree of how complex it is. We have conducted an evaluation
of the performance of these two novel features on EEG data corresponding
to motor-imagery. We also compared them to the most successful features
used in the BCI field, namely the Band-Power features. We evaluated
these three kinds of features and their combinations on EEG signals
from 13 subjects. Results obtained show that our novel features can
lead to BCI designs with improved classification performance, notably
when using and combining the three kinds of feature (band-power, multifractal
cumulants, predictive complexity) together.
\end{abstract}

\section{Introduction}

Brain-Computer Interfaces (BCI) are communication systems that enable
users to send commands to a computer by using only their brain activity
\cite{Wolpaw02}, this activity being generally measured using ElectroEncephaloGraphy
(EEG). Most EEG-based BCI are designed around a pattern recognition
approach: In a first step features describing the relevant information
embedded in the EEG signals are extracted \cite{Bashashati07}. They
are then feed into a classifier which identifies the class of the
mental state from these features \cite{reviewClassifEEG}. Therefore,
the efficiency of a BCI, in terms of recognition rate, depends mostly
on the choice of appropriate features and classifiers. Despite the
large number of features that have been explored to design BCI \cite{Bashashati07},
the performances of current EEG-based BCI are still not satisfactory,
and the BCI community has stressed the need to further explore and
design alternative features \cite{McFarland06}.

In this paper, we focus on feature extraction from EEG signals for
the design of BCI based on motor imagery (MI) \cite{Pfurtscheller01}.
MI corresponds to the imagination of limb movements (e.g., hand or
feet) without actual output. It is known to trigger brain signals
variations in specific frequency bands that can be used to drive a
BCI \cite{Pfurtscheller01}. The contribution of this paper is two-fold.
First, we introduce two new features for MI classification in EEG-based
BCI. These two new features are based on: 1. multifractal cumulants
\cite{EmpiricalMFA_WendtAbryJaffard} and 2. predictive complexity
\cite{DecisionalStates}. Second, we perform systematic comparisons
and analysis of the performance of these two new features, together
with the most successful feature for motor-imagery based-BCI, namely,
band-power feature \cite{spectralApproaches}.

The first new feature, namely, multifractal cumulants, can be seen
as a statistic on inter-frequency band relations. This is particularly
relevant for BCI as this information is generally ignored in current
motor imagery-based BCI designs, mostly based on the sole power in
different frequency bands. It should be mentioned that a preliminary
study of this kind of feature has been presented in a conference paper
\cite{MFAForBCI}. The second new feature is based on the statistical
complexity and predictive properties of the time series \cite{DecisionalStates}.
The information (quantified in bits) that is extracted this way measures
how difficult it is to make an optimal prediction based on past information.
It is null both for totally ordered and totally random systems, and
increases in between. It has already been applied to single simulated
neurons \cite{QuantLearnRecSpikNeurons} and a related form was applied
to measure synchronisation in the brain \cite{InfoCoherenceNNet}.
The assumption is that performing a mental task (e.g., motor imagery)
makes the EEG signal either more or less predictable, which can be
detected by a classifier when quantified by this second new feature.

The remainder of this paper is organised as follows: Sections \ref{sub:Multifractal-analysis}
and \ref{sec:Predictive-complexity-measure} present the two new features
that we propose. Section \ref{sec:Results} presents the experimental
evaluation, including the data sets that were used. The results are
then discussed. Section \ref{conclusion} concludes this study.

The code used for all the experiments in this paper is provided as
Free/Libre software. The data used is available online and all the
presented results are reproducible independently.

\section{Multifractal cumulants\label{sub:Multifractal-analysis}}

The multifractal formalism is described in details in \cite{EmpiricalMFA_WendtAbryJaffard,MFA_formalism_MuzyBacryArneodo}.
This section presents a short overview for the needs of this document.

Intuitively, the multifractal cumulants of the signal capture a signature
of inter-band relations (see below). This contrasts to the power in
each frequency band that is generally used. As shown in \cite{MFAForBCI}
the multifractal spectrum can in itself be used for EEG classification.
When considering multifractal in addition to power band feature vectors,
the resulting combination may improve the classification accuracy.

The method we chose for extracting the multifractal spectrum is a
discrete wavelet transform of the signal, out of which we extract
the wavelet leader coefficients \cite{WaveletLeaders}. Following
the directions of \cite{EmpiricalMFA_WendtAbryJaffard} we then use
the cumulants of the leaders as the features for classification, unlike
what we previously did in \cite{MFAForBCI}.

Let $x(t)$ be the signal to analyse. One view on multifractal analysis
\cite{mfa_asset_return} is to relate the statistical properties of
$x(t)$ and of a scaled version of it $x(at)$. In terms of frequency
analysis, that scaling in time corresponds to a frequency shift. Hence,
another view of the multifractal cumulants feature is that they characterise
some form of inter-frequency information, as mentioned in the introduction
of this section.

More precisely:
\begin{itemize}
\item The signal $x(t)$ is decomposed using a Discrete Wavelet Transform
to get the wavelet coefficients $w(s,t_{s})$ at each dyadic scale
$s$ and time interval $t_{s}$.
\item The wavelet leaders at each scale $s$ are then extracted by computing
the maxima of the wavelet coefficients over all samples involved for
computing $w(s,t_{s}-1)$, $w(s,t_{s})$ and $w(s,t_{s}+1)$ (including
lower scales) \cite{WaveletLeaders}.
\item Instead of performing a Legendre transform, or a direct Holder exponent
density estimation as in \cite{MFAForBCI}, we use here the recent
technique introduced by \cite{EmpiricalMFA_WendtAbryJaffard} and
compute the wavelet leader cumulants of orders 1 to 5. As noted in
\cite{EmpiricalMFA_WendtAbryJaffard} the first few cumulants already
contain most of the information useful in practice for characterising
the distribution of the Holder exponents. For a classification task
this information can now be exploited in a more condensed form.
\item The 5 first cumulants are computed for the leaders at each scale $s$.
Considering there is at most $L$ levels of wavelet transform in a
signal of size between $2^{L}$ and $2^{L+1}$, we get a total of
$5\times L$ cumulants for the signal, that progressively encompass
more and more frequency bands as the scale increases. These $5\times L$
cumulants per channel are used as the feature vector.
\end{itemize}
This method can be quite sensitive to the presence of electromagnetic
interferences at 50 Hz. We thus pre-filter the signals as described
in section \ref{sub:Data-sets} before proceeding to the multifractal
cumulants estimation.

\section{Predictive complexity measure\label{sec:Predictive-complexity-measure}}

This paper introduces for the first time the predictive complexity
measure of \cite{DecisionalStates} in the context of EEG classification.

The intuitive idea behind this feature is to quantify the amount of
information that is necessary to retain from the past of the series
in order to be able to predict optimally the future of the series
(\cite{ShaliziThesis}, and see below for the optimality criterion).

We had indication from related previous works that the feature could
be relevant for EEGs:
\begin{itemize}
\item At the level of neurons: statistical complexity was used to describe
the computational structure of spike trains \cite{spikeTrainsStruct}.
It was also shown to decrease while the neurons are learning in an
artificial spiking neural network \cite{QuantLearnRecSpikNeurons}.
\item At the level of the brain: information coherence and synchronisation
mechanism between communities of neurons are presented in \cite{InfoCoherenceNNet},
relying on related techniques.
\end{itemize}

\subsection{\label{sub:Decisional-states}Decisional states and the corresponding
complexity measure}

Informally, the idea behind the decisional states is to construct
a Markovian automaton \cite{epsilon-machine,ShaliziThesis} whose
states correspond to taking the same decisions \cite{DecisionalStates},
according to a user-defined utility function. These decisions are
those that one can take based on predictions of the future and their
expected utility.

The complexity of the series is then computed as the mutual information
between the internal states of the Markovian automaton, and the series
itself. The complexity is null for a very regular series, for example
a constant series or a series where we always take the same decision:
there is only a single state in the automaton. Similarly the complexity
of a completely random series is also null: it can be modelled by
successive independent draws from a fixed probability distribution,
whose expected utility we take to make our decision. This leads again
to a single Markovian automaton state, hence a null complexity. The
complexity measure increases only for more complicated series with
many internal states (i.e. many distinct probability distributions
of what happens next, depending on what previously happened, leading
to different decisions).

Presumably when the EEG corresponds or not to some functional activity,
the complexity of the series should change. The idea is to plug machine
learning techniques for monitoring that change.

\subsubsection*{Formal description}

Formally, let $\left(s_{t}\right)$ be a time series, with $t$ the
time index. Let $s_{t}^{-\infty}=\left(s_{u}\right)_{u\leq t}$ and
$s_{t}^{+\infty}=\left(s_{u}\right)_{t<u}$ be the past and future
histories at time $t$. In practice and for real measures, the time
range is finite: $0\leq t\leq T$. Similarly we measure the past and
future histories with finite horizons: $s_{t}^{-h}=\left(s_{u}\right)_{t-h\leq u\leq t}$
and $s_{t}^{+k}=\left(s_{u}\right)_{t<u\leq t+k}$.

Let us now consider predicting the future from the past statistically:
we seek to determine $P\left(s_{t}^{+k}|s_{t}^{-h}\right)$ at each
time $t$ (possibly with $h$ or $k$ infinite). Assuming the system
is conditionally stationary, that distribution does not change through
time: the same causes produce the same consequences. Let then $S^{-h}=\left\{ s_{t}^{-h}\right\} _{\forall t}$
and $S^{+k}=\left\{ s_{t}^{+k}\right\} _{\forall t}$ the sets of
all past and future histories. We will drop the time indices from
now on to indicate the time shift invariance. 

The causal states $\zeta$ are defined as the equivalence classes
of past histories with the same conditional distribution of futures:
$\zeta\left(s^{-h}\right)=\left\{ x\in S^{-h}:\, P\left(s^{+k}|x\right)=P\left(s^{+k}|s^{-h}\right)\right\} =\left\{ x:\, x\,\overset{c}{\equiv}s^{-h}\right\} $.
Knowing the causal state at the current time is the minimal information
needed for making optimal predictions \cite{ShaliziThesis} using
the full conditional probability distribution.

In the discrete case the series are strings of symbols drawn from
an alphabet $\mathcal{A}$. Each time step implies a symbol transition,
which possibly leads to a different causal state. The corresponding
automaton is called the ε-machine \cite{epsilon-machine}.

Let us now introduce a utility function $u:\,\left(S^{+k}\right)^{2}\mapsto\mathbb{R}$,
such that $u\left(r^{+k},s^{+k}\right)$ quantifies the gain (positive)
or loss (negative) when the user relied on the prediction $r^{+k}$
while $s^{+k}$ actually happened. We can now define an expected utility:
$\mathbb{U}\left[r^{+k}|s^{-h}\right]=\mathbb{E}_{s^{+k}\in S^{+k}}\left[u\left(r^{+k},s^{+k}\right)|s^{-h}\right]$,
quantifying what utility can be expected on average when choosing
the prediction $r^{+k}$ for the current system state $s^{-h}$. The
set of optimal predictions, realising the maximal expected utility,
can now be defined as $Y\left(s^{-h}\right)=\textrm{argmax}_{r^{+k}\in S^{+k}}\mathbb{U}\left[r^{+k}|s^{-h}\right]$.

The following equivalence relations $\overset{p}{\equiv}$, $\overset{u}{\equiv}$,
and $\overset{d}{\equiv}$ naturally extend the causal state equivalence
relation $\overset{c}{\equiv}$, taking into account the utility function:
\begin{itemize}
\item $r^{-h}\overset{p}{\equiv}s^{-h}$ when $Y\left(r^{-h}\right)=Y\left(s^{-h}\right)$,
with the corresponding \emph{iso-prediction sets} as equivalence classes.
All past histories within the same class lead to choosing the same
predictions, even though the expected utility may change from one
past history to the other.
\item $r^{-h}\overset{u}{\equiv}s^{-h}$ when $\max_{r^{+k}\in S^{+k}}\mathbb{U}\left[r^{+k}|r^{-h}\right]=\max_{r^{+k}\in S^{+k}}\mathbb{U}\left[r^{+k}|s^{-h}\right]$,
with the corresponding \emph{iso-utility sets} as equivalence classes.
All past histories within the same class lead to the same maximum
expected utility, even though the optimal predictions to choose for
reaching this utility are not specified.
\item $r^{-h}\overset{d}{\equiv}s^{-h}$ when $r^{-h}\overset{p}{\equiv}s^{-h}$
and $r^{-h}\overset{u}{\equiv}s^{-h}$. The intersection of the above
iso-utility and iso-prediction sets define the \emph{decisional states}:
$\Psi(s^{-k})=\left\{ x\in S^{-h}:\, x\overset{p}{\equiv}s^{-h},\, x\overset{u}{\equiv}s^{-h}\right\} $.
We assume that when both the maximal expected utility and the optimal
predictions are the same, the user will reach the same decisions.
In other words, the utility function encodes all the user needs to
know to reach a decision.
\end{itemize}
It can be easily shown \cite{DecisionalStates} that the causal states
sub-partition the decisional states. That is to say, the causal states
have lost their minimality property due to the fact we are not interested
in the full conditional distribution of futures but only in the optimal
decisions with respect to a user-defined utility function.

The causal states are an intrinsic property of the data set. The mutual
information $M_{c}=I(s^{-h};\zeta\left(s^{-h}\right))$ between the
causal $\zeta\left(s^{-h}\right)$ state and the series of $s^{-h}$
defines an intrinsic measure, the statistical complexity \cite{epsilon-machine},
quantifying how hard it is to get the conditional distribution of
futures from the current observed past.

The decisional states correspond to the structure implied by the user
utility function on top of the causal states. As for the causal states,
knowing the decisional state at the current time is the information
needed for making optimal predictions maximising the user-defined
utility function. The mutual information $M_{d}=I(s^{-h};\Psi\left(s^{-h}\right))$
similarly defines a complexity measure for how hard it is to make
these predictions, called decisional complexity by analogy with the
statistical complexity.

\subsection{\label{sub:Complexity on EEG}Application to EEG data}

In the present study the chosen utility function is the negative sum
of square error: $u\left(r^{+k},s^{+k}\right)=-\left\Vert r^{+k}-s^{+k}\right\Vert ^{2}$.

Each EEG series is split in time windows of $h+k$ samples, defining
each a $\left(s^{-h},s^{+k}\right)$ pair. These observations are
fed in the reconstruction algorithm presented in \cite{DecisionalStates}.
That algorithm estimates the joint probabilities $p\left(s^{-h},s^{+k}\right)$
using a kernel density estimation with Gaussian kernels. The conditional
probabilities are then computed by integration over $S^{+k}$: $p\left(s^{+k}|s^{-h}\right)=p\left(s^{-h},s^{+k}\right)/\int_{\sigma^{+k}\in S^{+k}}p\left(s^{-h},\sigma^{+k}\right)$.
These are then clustered into causal state estimates $\zeta\left(s^{-h}\right)=\left\{ x\in S^{-h}:\, p\left(s^{+k}|x\right)=p\left(s^{+k}|s^{-h}\right)\right\} $
using connected components up to a fixed Bhattacharyya distance threshold
(chosen to be $0.05$) between the conditional distributions. The
utility function is applied on top of the aggregated causal states
distributions $p\left(s^{+k}|\zeta\right)=\mbox{avg}_{s^{-h}\in\lyxmathsym{ζ}}p\left(s^{+k}|s^{-h}\right)$
in order to get the expected utilities $\mathbb{U}\left[r^{+k}|\zeta\right]$.
Finally, the causal states are themselves clustered into iso-utility
and iso-prediction sets, which are intersected to get the decisional
states $\lyxmathsym{ω}$.

The decisional complexity is the feature that is extracted from the
EEG series. We thus get one feature per electrode.

\section{Evaluation\label{sec:Results}}

\subsection{\label{sub:Data-sets}Data sets}

Four data sets for a total of 13 subjects were used for evaluation
in this study. Data for 12 subjects come from the international BCI
competitions%
\footnote{http://www.bbci.de/competition/%
} II, III, and IV \cite{Blankertz04}\cite{Blankertz06}. The data
for the last subject was acquired at INRIA (French National Research
Institute on Computer Science and Control) Rennes-Bretagne Atlantique,
using the OpenViBE software platform \cite{Renard10}%
\footnote{http://openvibe.inria.fr/?q=datasets%
}.

\subsubsection{BCI competition II, data set III (1 subject)}

This data set was captured at the Department of Medical Informatics,
Institute for Biomedical Engineering, University of Technology Graz
\cite{BCICompetitonData}. The data contains 280 trials sampled at
128 Hz. During each trial, the subject is presented with a visual
cue indicating either left or right at random, and shall then imagine
a movement of the corresponding hand during 6s. There was 140 trials
of left class (left hand motor imagery) and 140 trials of the right
class (right hand motor imagery). EEG were recorded using the C3,
C4 and Cz electrodes, however, for the purpose of this evaluation,
we used only the C3 and C4 electrodes as recommended in \cite{Lemm04}.
More details about this data set can be found in \cite{Blankertz04}.

The data was already preprocessed by a band-pass filter between 0.5
and 30 Hz. Unfortunately, the nature of the filter is not specified,
and the DC component was not well removed. Since this interferes in
particular with the algorithm for estimating the series complexity,
a new filter is applied with the following characteristics: 
\begin{itemize}
\item FIR filter obtained using METEOR \cite{METEOR} 
\item Less than 1dB change in 4-30Hz with a linear phase response 
\item At least -50dB at 0 Hz (and above 40 Hz, though in the present case
the signal is already filtered) 
\item 1/4 sec delay 
\end{itemize}
We then used all the available filtered data over the feedback period
in order to extract the features.

\subsubsection{BCI competition III, data set IIIb (2 subjects)}

This data set was also captured at the Department of Medical Informatics,
Institute for Biomedical Engineering, University of Technology Graz
\cite{BCICompetitonData}. It originally consists of EEG signals from
3 subjects who performed left hand and right hand motor imagery. However,
for the purpose of this study, only subjects labeled S4 and X11 were
used. Indeed, EEG signals for subject O3VR were recorded using a different
protocol and the data file provided online contained erroneously duplicate
signals%
\footnote{See http://www.bbci.de/competition/iii/desc\_IIIb\_ps.html for details%
}. The experimental protocol for subjects S4 and X11 is similar to
the one used for the BCI competition II data set III. The data was
captured at a 125 Hz sampling rate using electrodes C3 and C4. More
details about this data set can be found in \cite{Blankertz06}. EEG
signals in this data set were already band-pass filtered in 0.5-30
Hz. As for the previous data set, an additional FIR filter with the
same characteristics as the previous one was applied, for the same
reasons. We also used all the available filtered data over the feedback
period in order to extract the features.

\subsubsection{BCI competition IV, data set IIb (9 subjects)}

This third data set was provided by the same team and follows the
same experimental protocol as the above two. However it contains occular
artifacts that interfere with the brain signals. Although this is
less convenient for the purpose of testing our two new features it
is also more realistic. We thus choosed to ignore the artifacts and
process as if the signals were clean EEGs, in order to assert the
robustness of our features to the occular artifacts.

Data coming from nine subjets is sampled at 250 Hz, and pre-processed
by a band-pass filter between 0.5 and 100 Hz with a notch at 50 Hz.
For the aforementioned reasons we had to re-filter the signals. We
choosed a FIR design with a band-pass between 6 and 30 Hz, with a
null at 0Hz to suppress the DC component and -50dB above 40 Hz.

We also used all the available filtered data over the feedback period
in order to extract the features.

\subsubsection{OpenViBE / INRIA data (1 subject)}

This data set was captured at INRIA Rennes-Bretagne Atlantique using
the OpenViBE free and open-source software \cite{Renard10}. This
data set comprises EEG signals from one subject who performed left
hand and right hand motor imagery. 560 trials of motor imagery (280
trials per class) were recorded over a 2 week period. Data were collected
using the same experimental protocol as the one used for the BCI competition
data, i.e., the Graz BCI protocol \cite{Pfurtscheller01}. Half of
the trials (randomly selected from all experiments over time) is used
for training, the remaining half for testing.

EEG data was sampled at 512 Hz and recorded using the following electrodes:
C3, C4, FC3, FC4, C5, C1, C2, C6, CP3, CP4, with a nose reference
electrode. The use of such electrodes enables us to apply a discrete
Laplacian spatial filter \cite{LaplacianFiltering} over C3 and C4
in order to obtain better signals, as recommended in \cite{McFarland97}.

The data was preprocessed by a FIR filter (designed with METEOR \cite{METEOR})
with the following characteristics:
\begin{itemize}
\item Less than 1dB change in 4-30Hz, linear phase response in this range.
\item At least -50dB at 0 Hz and above 50 Hz.
\item Null responses at 50 Hz and all harmonics.
\item 1/4 sec delay.
\end{itemize}
We then used all the available filtered data over the feedback period
in order to extract the features.

\subsection{Results obtained with all features and their combinations}

The goal of this section is to show what results can be obtained with
each feature considered in isolation, and the effect of combining
them. The hypothesis we wish to test is that each feature extracts
a different information from the signal, and thus that combining them
can improve the classification accuracy.

\subsubsection{Features}

For the experiments below, the following methodology was applied:
\begin{itemize}
\item The power in frequency bands was extracted for each subject. As was
shown by \cite{spectralApproaches} and in our own forthcoming study
\cite{powcomp} estimating this information from data is sensitive
to choice of the time-frequency decomposition algorithm. We extracted
the power information in each frequency band using a Morlet Wavelet
which support is determined by cross-validation. All bands were then
kept between 4 and 30 hertz, for each of the C3 and C4 electrode:
this leads to a 52-dimensional feature vector.
\item The multifractal cumulants feature was extracted according to the
method described in Section \ref{sub:Multifractal-analysis}. Cross-validation
on the training set was used to determine the wavelet support and
the number of decomposition levels to retain. Since we have 5 cumulants
per level per electrode, this leads to a $10*N$-dimensional vector
with $N$ the number of retained levels.
\item Parameters for estimating the predictive complexity feature were also
determined by maximising the cross-validation performance on the training
set: the number of points to retain from the past and the future,
the sub-sampling factor for the series, the kernel size used for estimating
the conditional probability distributions, and whether to work on
the raw series or the first differences. Cross-validation selected
1 point in the future, 5 points from the past with a sub-sampling
factor of 8 for the OpenViBE subject, and 6 points from the past with
a sub-sampling factor of 2 for the BCI II and III subjets. These sub-sampling
parameters correspond to having a sampling frequency of 62.5 Hz for
the S4 and X11 subjects and 64 Hz for the others, which thanks to
Nyquist theorem matches the filtering operation that was performed
on the signals: cross-validation selected the most economical sub-sampling
parameter that still captures the remaining frequencies in the signal
between 4 and 30 Hz. We thus saved computational time for the BCI
IV subjets by fixing the subsampling parameters and cross-validating
only the kernel width. All sub-sampled signals (ex: all 8 possible
series for a sub-sampling by factor 8) were kept for building the
statistics in the complexity computations.
\end{itemize}
Therefore the feature vectors include:
\begin{itemize}
\item 52 power features for frequencies between 4 and 30 Hz. (26 for each
C3 / C4 electrode)
\item 30, 40, or 50 features for multifractal cumulants.
\item 2 features for the predictive complexity (one for each electrode).
\end{itemize}
As a side remark we shall point out that cross-validating all the
above parameters required a quite consequent computational power,
which was obtained using the BOINC distributed computing infrastructure
\cite{BOINC}. Nevertheless, once the optimal parameters were determined
off-line by cross-validation, the computational requirements for extracting
the features are nothing exceptional: the features can be computed
in real time on a standard PC.

\subsubsection{Classifier and combination rule}

In order to classify the extracted features, we used a Linear Discriminant
Analysis (LDA), one of the most popular and effi{}cient classifi{}er
for EEG-based BCI \cite{reviewClassifEEG}.

Each series was classified independently into either the {}``left
hand'' or {}``right hand'' class of the motor imagery task using
each feature independently. The percentage of correct classifications
on the test set is reported for each subjet in the first 3 colums
of Table \ref{tab:results}. 

We also evaluated the accuracy obtained when combining these three
features together. To do so, we first estimate the global accuracy
of each feature alone by using the Fisher discriminant ratio $F$
on the training data set, hence we get $F^{\textrm{pow}}$, $F^{\textrm{mfa}}$,
$F^{\textrm{cpx}}$ for the three features Band-Power, Multifractal
cumulants, and Complexity. In order to combine the results we gather
the outputs of the 3 linear classifi{}ers trained independently on
each feature by performing a simple arithmetic average for each instance
$P_{i}^{\textrm{combi}}=F^{\textrm{pow}}*P_{i}^{\textrm{pow}}+F^{\textrm{mfa}}*P_{i}^{\textrm{mfa}}+F^{\textrm{cpx}}*P_{i}^{\textrm{cpx}}$
where each $P_{i}$ is a prediction (classifier output) for the instance
number $i$. That value is then used for classifying the instance
in the results reported in the fourth column of table \ref{tab:results}.

\subsubsection{Results}

\selectlanguage{english}%
\begin{table}
\begin{raggedright}
\begin{tabular}{|l|l|l|l|l|l|}
\hline 
\selectlanguage{british}%
Subjet\selectlanguage{english}
 & \selectlanguage{british}%
Band-Power\selectlanguage{english}
 & \selectlanguage{british}%
Multifractal\selectlanguage{english}
 & \selectlanguage{british}%
Complexity\selectlanguage{english}
 & \selectlanguage{british}%
Combination\selectlanguage{english}
 & \selectlanguage{british}%
Gain\selectlanguage{english}
\tabularnewline
\hline 
\selectlanguage{british}%
bci2\selectlanguage{english}
 & 77.1 & 80.7 & 77.9 & \selectlanguage{british}%
\emph{\cellcolor{green}}\foreignlanguage{english}{80.7}\selectlanguage{english}
 & 3.57\tabularnewline
\hline 
\selectlanguage{british}%
s4\selectlanguage{english}
 & 81.5 & 74.4 & 65.0 & \selectlanguage{british}%
\emph{\cellcolor{cyan}}\foreignlanguage{english}{81.7}\selectlanguage{english}
 & 0.19\tabularnewline
\hline 
\selectlanguage{british}%
x11\selectlanguage{english}
 & 80.4 & 68.7 & 70.2 & \selectlanguage{british}%
\emph{\cellcolor{cyan}}\foreignlanguage{english}{80.9}\selectlanguage{english}
 & 0.56\tabularnewline
\hline 
\selectlanguage{british}%
OpenVibe\selectlanguage{english}
 & 92.9 & 85.7 & 76.1 & \selectlanguage{british}%
\emph{\cellcolor{cyan}}\foreignlanguage{english}{93.2}\selectlanguage{english}
 & 0.36\tabularnewline
\hline 
\selectlanguage{british}%
bci4s1\selectlanguage{english}
 & 77.5 & 65.9 & 55.9 & \selectlanguage{british}%
\emph{\cellcolor{orange}}\foreignlanguage{english}{74.4}\selectlanguage{english}
 & -3.12\tabularnewline
\hline 
\selectlanguage{british}%
bci4s2\selectlanguage{english}
 & 56.4 & 57.5 & 56.1 & \selectlanguage{british}%
\emph{\cellcolor{cyan}}\foreignlanguage{english}{56.1}\selectlanguage{english}
 & -0.36\tabularnewline
\hline 
\selectlanguage{british}%
bci4s3\selectlanguage{english}
 & 51.9 & 51.2 & 54.1 & \selectlanguage{british}%
\emph{\cellcolor{cyan}}\foreignlanguage{english}{52.2}\selectlanguage{english}
 & 0.31\tabularnewline
\hline 
\selectlanguage{british}%
bci4s4\selectlanguage{english}
 & 93.4 & 91.2 & 90.3 & \selectlanguage{british}%
\emph{\cellcolor{green}}\foreignlanguage{english}{95.6}\selectlanguage{english}
 & 2.19\tabularnewline
\hline 
\selectlanguage{british}%
bci4s5\selectlanguage{english}
 & 96.9 & 88.8 & 74.1 & \selectlanguage{british}%
\emph{\cellcolor{cyan}}\foreignlanguage{english}{95.9}\selectlanguage{english}
 & -0.94\tabularnewline
\hline 
\selectlanguage{british}%
bci4s6\selectlanguage{english}
 & 87.8 & 82.2 & 68.4 & \selectlanguage{british}%
\emph{\cellcolor{green}}\foreignlanguage{english}{89.1}\selectlanguage{english}
 & 1.25\tabularnewline
\hline 
\selectlanguage{british}%
bci4s7\selectlanguage{english}
 & 70.6 & 68.1 & 70.3 & \selectlanguage{british}%
\emph{\cellcolor{green}}\foreignlanguage{english}{72.2}\selectlanguage{english}
 & 1.56\tabularnewline
\hline 
\selectlanguage{british}%
bci4s8\selectlanguage{english}
 & 80.0 & 88.4 & 90.6 & \selectlanguage{british}%
\emph{\cellcolor{green}}\foreignlanguage{english}{89.1}\selectlanguage{english}
 & 9.06\tabularnewline
\hline 
\selectlanguage{british}%
bci4s9\selectlanguage{english}
 & 78.8 & 83.1 & 78.8 & \selectlanguage{british}%
\emph{\cellcolor{green}}\foreignlanguage{english}{82.8}\selectlanguage{english}
 & 4.06\tabularnewline
\hline
\end{tabular}
\par\end{raggedright}

\selectlanguage{british}%
{\small Legend: Green when the classification accuracy of the combination
is higher by at least 1\% than when using only the usual Band-Power
feature, blue when the change in performance with the combination
is between -1\% and 1\%, orange when the combination is worse.}{\small \par}

\selectlanguage{english}%
\caption{\selectlanguage{british}%
\label{tab:results}Classification accuracies over the test sets for
each subject and feature\selectlanguage{english}
}

\end{table}

\selectlanguage{british}%
The multifractal cumulants and the predictive complexity features
allow by themselves to classify the BCI data sets with a good accuracy.
This, in itself, is a contribution of this paper. Although the accuracy
obtained using either the Multifractal Cumulants or the Predictive
Complexity is slightly lower on average that when using only the Power
Bands feature (sometimes higher), we have extracted some information
from the signal that is adequate for classification.

Results also showed that combining all features is better than using
the Power Bands alone for 6 subjects, on the same order for 6 subjects,
and with a negative impact for only one subject. A gain of $1.4$\%
accuracy was obtained on average, with much higher values for some
subjects (see table \ref{tab:results}). This not only confirms that
some different information was extracted from the signal, but also
stresses the usefulness of our new features for BCI classification
tasks.

\section{Conclusion\label{conclusion}}

This study has introduced two new features for Brain-Computer Interface
design: multifractal cumulants and predictive complexity. The information
contained in the multifractal cumulants feature corresponds to a relation
between frequency bands, rather than the power extracted in each band.
The complexity feature measures the difficulty to predict the future
of the EEG signals based on their past.

Interestingly enough, our results showed that the two new features,
i.e., multifractal cumulants and predictive complexity measure, could
indeed be used by themselves to discriminate between different motor
imagery mental states as measured by EEG. This is especially interesting
as the complexity feature only adds one scalar value per electrode,
compared to the other features we considered (ex: one dimension per
frequency band in the band-power case).

Moreover, our results showed that when combining these two features
together with band-power features, the resulting BCI could reach a
better classification accuracy than when using the usual band-power
feature alone. Thus this suggests that these new features are a good
complement to currently used features for BCI design and that they
can lead to improved BCI design. Therefore we would recommend BCI
designers to consider these two features as additional features in
the conception of a BCI system, in order to obtain better performance.

Future work might involve the exploration of novel ways to combine
the features and feature selection, as well as the application of
these features to BCI tasks other than motor imagery. Work is also
needed on the design of novel algorithms including physiologically
relevant error functions for EEG signal predictions for the complexity
feature.

\part*{Appendix: Source code, data, and web information}

All the results in this study are reproducible independently. The
code for the experiments is provided as Free/Libre software on the
main author web site:
\begin{itemize}
\item http://nicolas.brodu.numerimoire.net/en/recherche/publications/
\end{itemize}
The data that was used can be downloaded on the BCI competitions website
and on the OpenViBE project page:
\begin{itemize}
\item http://www.bbci.de/competition/
\item http://openvibe.inria.fr/?q=datasets
\end{itemize}
\bibliographystyle{plain}
\bibliography{bci_mfac_pc}

\end{document}